\documentclass[12pt]{article}

\usepackage[left=2.8cm,right=2.8cm,top=2.8cm,bottom=2.8cm]{geometry}

\usepackage[colorlinks=true,linkcolor=black,citecolor=black,urlcolor=blue,filecolor=black]{hyperref}

 \csname
@addtoreset\endcsname{equation}{section}

\usepackage[utf8]{inputenc}
\usepackage{amsmath}
\usepackage{amsfonts}
\usepackage{amssymb}
\usepackage{stmaryrd}
\usepackage{mathtools}

\usepackage[sort]{cite}
\usepackage[usenames,table]{xcolor}
\usepackage{dsfont}
\usepackage{mathrsfs}

\def\cE{{\cal E}}
\def\cF{{\cal F}}

\def\cH{{\cal H}}

\def\cL{{\cal L}}

\def\cR{{\cal R}}

\def\cT{{\cal T}}

\def\g{\gamma} 

\def\d{\delta} 

\def\e{\epsilon}

\def\h{\eta}

\def\la{\lambda}
\def\La{\Lambda}
\def\m{\mu}
\def\n{\nu}
\def\p{\pi}

\def\x{\xi}

\def\s{\sigma}

\def\t{\tau}

\def\O{\Omega}

\newcommand{\be}{\begin{equation}}
\newcommand{\ee}{\end{equation}}

\def\pr{\partial}

\def\nn{\nonumber}

\def\lapse{N}
\def\frame{\mathfrak{e}}

\begin{document}

\vspace{30pt}

\begin{center}


{\Large\sc Magnetic Carrollian gravity\\[10pt] from the Carroll algebra}

\vspace{-5pt}
\par\noindent\rule{350pt}{0.4pt}


\vspace{20pt}
{\sc 
Andrea~Campoleoni${}^{\; a,}$\footnote{Research Associate of the Fund for Scientific Research – FNRS, Belgium.},
Marc~Henneaux${}^{\; b, c}$,
Simon Pekar${}^{\; a,}$\footnote{FRIA grantee of the Fund for Scientific Research – FNRS, Belgium.},
Alfredo~Pérez${}^{\; d, e}$ and
Patricio~Salgado-Rebolledo${}^{\; f}$
}

\vspace{8pt}
${}^a${\it\small
Service de Physique de l'Univers, Champs et Gravitation,\\
Universit{\'e} de Mons -- UMONS,
20 place du Parc, 7000 Mons, Belgium}
\vspace{4pt}

${}^b${\it\small 
Université Libre de Bruxelles and International Solvay Institutes,\\ 
ULB -- Campus Plaine CP231,
1050 Brussels, Belgium}
\vspace{4pt}

${}^c${\it\small 
Coll\`ege de France, 11 place Marcelin Berthelot, 75005 Paris, France}
\vspace{4pt}

${}^d${\it\small
Centro de Estudios Cient\'ificos (CECs), Avenida Arturo Prat 514, Valdivia, Chile}
\vspace{4pt}

${}^e${\it\small
Facultad de Ingeniería, Arquitectura y Diseño, Universidad San Sebastián, sede Valdivia, General Lagos 1163, Valdivia 5110693, Chile}
\vspace{4pt}

${}^f${\it\small
Department of Theoretical Physics, Wroc\l aw University of Science and Technology,\\
50-370 Wroc\l aw, Poland}
\vspace{5pt}

{\tt\small 
\href{mailto:andrea.campoleoni@umons.ac.be}{andrea.campoleoni@umons.ac.be},
\href{mailto:marc.henneaux@ulb.be}{marc.henneaux@ulb.be},\\
\href{mailto:simon.pekar@umons.ac.be}{simon.pekar@umons.ac.be},
\href{mailto:alfredo.perez@uss.cl}{alfredo.perez@uss.cl},
\href{mailto:patricio.salgado-rebolledo@pwr.edu.pl}{patricio.salgado-rebolledo@pwr.edu.pl}
}

\vspace{30pt} {\sc\large Abstract}  \end{center}

\noindent
We explicitly establish the equivalence between the magnetic Carrollian limit of Einstein gravity defined through the Hamiltonian formalism and the Carrollian theory of gravity defined through a gauging of the Carroll algebra along the lines of standard Poincar\'e (or (A)dS) gaugings.

\newpage



\tableofcontents

\section{Introduction}

The interest of the Carrollian (or ultra-relativistic) contraction of the Poincar\'e algebra \cite{Levy:1965, SenGupta:1966} in the context of gravity first appeared in the study of the ``strong coupling''  \cite{Isham:1975ur} or zero signature \cite{Teitelboim:1978wv, Teitelboim:1981ua} limit of Einstein's theory. Initially expressed in Hamiltonian terms, the covariant formulation of that limit was worked out in \cite{Henneaux:1979vn}, where the underlying Carrollian geometry was identified and constructed. As realized more recently \cite{Henneaux:2021yzg}, the limit appearing in \cite{Isham:1975ur, Teitelboim:1978wv, Henneaux:1979vn, Teitelboim:1981ua} is the ``electric'' Carrollian contraction of Einstein gravity.  There is another Carrollian contraction, called ``magnetic'' in \cite{Henneaux:2021yzg}, which is also easily obtained in the Hamiltonian formalism. Conserved charges and asymptotic symmetries for both ``electric'' and ``magnetic'' Carrollian gravity were then studied in \cite{Perez:2021abf,Perez:2022jpr,Fuentealba:2022gdx}.
Aside from the original motivation, 
Carrollian field theories and gravity recently received a renewed interest due to their relation with the physics taking place at the null boundaries of asymptotically-flat spacetimes in general relativity, see e.g.~\cite{Duval:2014uva, Bagchi:2016bcd, Ciambelli:2018wre,  Donnay:2022aba}.

Specifically, the magnetic Carrollian limit of Einstein gravity on which we focus in this note can be derived as follows.
One starts from the canonical action in $D+1$ spacetime dimensions,
\be \label{eq:relativistic_ADM}
I=\int dt\,d^{D}x \left(\pi^{ij}\dot{g}_{ij}-\lapse \mathcal{H}_{\perp}-N^{i}\mathcal{H}_{i}\right) .
\ee
The Hamiltonian constraints take the form
\be
\mathcal{H}_{\perp} = \mathcal{H}_M +\mathcal{H}_E \, ,
\qquad 
\mathcal{H}_{i} = -\,2\,\nabla_{\!j}\pi_{i}{}^{j}\,,
\ee
where
\be\label{eq:electricandmagneticH}
\mathcal{H}_M = -\frac{\sqrt{g}}{16\pi G_M}\left(R-2\Lambda\right) ,
\qquad
\mathcal{H}_E = \frac{16\pi G_M c^2}{\sqrt{g} } \left(\pi^{ij}\pi_{ij}-\frac{1}{D-1}\, \pi^{2}\right) , 
\ee
with $G_M = c^{-4} G_N$ and $G_N$ denoting Newton's constant. One can then take the limit $c \to 0$, effectively dropping the term $\mathcal{H}_E$ in the Hamiltonian density, which is subleading in that expansion \cite{Henneaux:2021yzg}. One thus obtains the action of magnetic Carrollian gravity:
\be \label{eq:magnetic_action}
I_M=\int dt\,d^{D}x \left(\pi^{ij}\dot{h}_{ij}-\lapse \mathcal{H}_{M}-N^{i}\mathcal{H}_{i}\right) .
\ee
Signals of the existence of a magnetic Carrollian limit of general relativity were also pointed out in \cite{deBoer:2021jej} by implementing the limit $c \to 0$ on specific solutions. The magnetic theory was also recovered in \cite{Hansen:2021fxi} at the next-to-leading order in a Carrollian expansion of the covariant Einstein-Hilbert action, developed along the lines pioneered in \cite{Dautcourt:1997hb}. In the Hamiltonian setup, the electric Carrollian limit is obtained instead by rescaling the fields and Newton's constant so as to keep in the limit $c \to 0$ the term $\cH_{E}$ in the Hamiltonian density while dropping $\cH_{M}$, without affecting neither the kinetic term nor $\cH_i$.

The key dynamical feature that distinguishes the magnetic contraction from the electric one is that the momentum $\pi^{ij}$ conjugate to the metric cannot be eliminated using its own equation of motion in the magnetic case, while it can in the electric case. In the electric theory, eliminating $\pi^{ij}$ leads to the second-order covariant action of \cite{Henneaux:1979vn}, while in the magnetic theory the equation of motion for $\pi^{ij}$ forces the Carrollian second fundamental form (or extrinsic curvature) to vanish. This property plays a significant role below.

There exist other approaches to Carrollian theories of gravity, which are based on the gauging of the Carroll algebra along the lines of standard Poincar\'e (or (A)dS) gaugings \cite{Hartong:2015xda, Bergshoeff:2017btm, Figueroa-OFarrill:2022mcy, Guerrieri:2021cdz}. A natural question is then to compare the Hamiltonian and gauging procedures.
The form of the equations of motion points towards the identification of the theory developed in \cite{Hartong:2015xda} with electric Carrollian gravity, and of that developed in \cite{Bergshoeff:2017btm} with magnetic Carrollian gravity. In this paper, we focus on the magnetic case and we prove that the actions presented in \cite{Bergshoeff:2017btm} and \cite{Henneaux:2021yzg} are indeed equivalent.

Our paper is organized as follows. In section~\ref{sec:gauging}, we introduce the quantities that appear in the gauging of the Carroll algebra and that give a Cartan description of Carrollian geometry.  In section~\ref{sec:gauging2}, we prove the equivalence of the first-order action of \cite{Bergshoeff:2017btm} with the Hamiltonian action of \cite{Henneaux:2021yzg}. In section~\ref{sec:limit}, we revisit this result starting from the relation between the first-order and Hamiltonian formulations of Einstein gravity and implementing the limit $c \to 0$ afterward. This allows us to show how the key features of the magnetic theory are already visible in general relativity provided that one solves only part of the torsion constraint. In Appendix~\ref{app:electric}, we discuss how one could modify this limiting procedure to obtain instead the electric theory, highlighting however a series of unsatisfactory features that are absent in the magnetic limit.

\section{Gauging of the Carroll algebra - Kinematics}\label{sec:gauging}

We review in this section Carrollian geometry from the point of view of Carrollian connections and ``solderings''.  A soldering attaches to the manifold the connection components associated with translations  so that they can be regarded as tangent vectors to the manifold (``local frames''). This identification endows the tangent spaces to the manifold with the structure of a flat Carroll spacetime, on which the homogeneous Carroll group acts.  This approach to Carrollian geometry is quite general and necessary for understanding the gauging procedure to be presented later; see also the reviews \cite{Herfray:2021qmp, Bergshoeff:2022eog} and references therein for related discussions. We develop in this section the concepts per se, without any reference to a limiting procedure that would regard the Carrollian structure as a contraction of the corresponding Poincar\'e one.  

\subsection{Carroll algebra}

The Carroll algebra \cite{Levy:1965, SenGupta:1966} is a contraction of the Poincar\'e algebra and its non-vanishing commutators are
\begin{subequations} \label{Carroll-algebra}
\begin{align}
[ J_{ab}, P_c ] & = \delta_{cb}P_a-\delta_{ca}P_b \, ,\\[5pt]
[ J_{ab},C_c] & = \delta_{cb}C_a-\delta_{ca}C_b \, , \\[5pt]
[ J_{ab},J_{cd} ] & = \delta_{ac}J_{db}+\delta_{bd}J_{ca}-\delta_{ad}J_{cb}-\delta_{bc}J_{da} \, , \\[5pt]
[C_a,P_b] & = \delta_{ab} H \, .
\end{align}
\end{subequations}
The Latin indices $a, b, \ldots$ from the beginning of the alphabet are internal indices taking spatial values $1, \ldots, D$. The corresponding capital letters $A, B, \ldots$  take the spacetime values $0,1,\ldots,D$.  The generators $H$, $P_{a}$, $C_{a}$ and $J_{ab}$ are associated with time translations, spatial translations, Carrollian boosts and spatial rotations, respectively.  Carrollian boosts and spatial rotations define the homogeneous Carroll subalgebra.

Carroll transformations leave both the degenerate metric
\be
(\zeta_{AB}) = \begin{pmatrix} 0 & 0 \\ 0 & \delta_{ab} \end{pmatrix} \label{eq:internal_metric}
\ee
and the vector 
\be
(n^A) = \begin{pmatrix} 1 \\ 0  \end{pmatrix} , \qquad \zeta_{AB}\, n^B= 0 \label{eq:internal_vector}
\ee
invariant. One can in fact define the Carroll group as the group of linear transformations that leave these objects invariant.  The degenerate metric enables one to lower meaningfully (i.e., in a Carroll-invariant way) the internal indices.
The above commutation relations can be written in a compact way as
\begin{subequations} \label{Carroll-algebra-compact}
\begin{align}
[ J_{AB}, P_C ] & = \zeta_{CB}P_A-\zeta_{CA}P_B \, ,\\[5pt]
[ J_{AB},J_{CD} ] & = \zeta_{AC}J_{DB}+\zeta_{BD}J_{CA}-\zeta_{AD}J_{CB}-\zeta_{BC}J_{DA} \, ,
\end{align}
\end{subequations}
with $P_0 =H$ and $J_{0b} = C_b = -J_{b0}$.

One can also include a cosmological constant $\Lambda=\frac{\sigma D(D-1)}{2\ell^2}$ by considering the Carroll (A)dS algebra, which is a contraction of the (A)dS algebra and includes the additional non-vanishing commutators
\be
[ H, P_a ]= -\frac{\sigma}{\ell^2}\, C_a \,, \qquad
[P_a, P_b ]= -\frac{\sigma}{\ell^2}\, J_{ab} \,,
\ee
or together as
\be
[P_A, P_B] = -\frac{\sigma}{\ell^2}\, J_{AB} \,,
\ee
where $\sigma=1$ ($-1$) corresponds to the dS (AdS) case.

\subsection{Carrollian connection -- Vielbein}

The first step in the ``gauging'' of a Lie algebra is to define a connection one-form taking values on that algebra, here the Carrollian one (or its (A)dS counterpart):
\be \label{connection}
A_{\mu} = \tau_\mu H+e_\mu{}^a P_{a}+\omega_\mu{}^a C_{a}+\frac{1}{2}\,\omega_\mu{}^{ab} J_{ab} \,,
\ee
where $\mu$ are $D+1$ spacetime indices.

While the (local) introduction of a connection one-form can be done for any algebra, the next step to be discussed uses the particular feature that some of the generators of the Carroll algebra are translations and can be identified with a basis of tangent (co)vectors to the manifold.
Specifically, we shall make throughout the non-degeneracy assumption that the one-forms $\tau_\mu$,  $e_\mu{}^a$ are linearly independent, so that the set of $D+1$ one-forms $\{\tau_\mu, e_\mu{}^a\}$ constitute a basis of the cotangent space (``soldering'').  This implies that the determinant 
\be
\mathcal E=\det\left(\tau_{\mu},e_{\mu}{}^{a}\right) = \frac{1}{D!}\, \epsilon_{a_1\cdots a_D}\epsilon^{\mu_0\cdots \mu_D}\tau_{\mu_0} e_{\mu_1}{}^{a_1}\cdots e_{\mu_D}{}^{a_D} \not= 0
\ee
does not vanish. The dual basis of the tangent space is denoted by $\{n^{\mu}, e^{\mu}{}_a \}$ and fulfills
\be
e_{\mu}{}^{a} e^{\mu}{}_{b} =\delta^a{}_b \,, \quad 
\tau_{\mu}n^{\mu} = 1 \,, \quad
n^{\mu}e_{\mu}{}^{a} = 0 \,,\quad
\tau_{\mu}e^{\mu}{}_{a} = 0 \,.
\ee
From these relations, one derives
\be
e_{\mu}{}^{a}e^{\nu}{}_{a} + \tau_{\mu} n^{\nu}=\delta_{\mu}{}^{\nu} \, ,
\ee
(right and left inverses coincide) and
\be
\det\left(n^{\mu},e^{\mu}{}_{a}\right) = \mathcal E^{-1} \, .
\ee

The basis of tangent vectors $\{n^{\mu}, e^{\mu}{}_a \}$ is called the ``vielbein'', or ``local frame'' while the one-forms $\{\tau_\mu, e_\mu{}^a\}$ are the ``inverse vielbein'', or ``dual local frame''.\footnote{We resort here to the notation which is customary in general relativity, but we stress that in the supergravity literature (and in \cite{Bergshoeff:2017btm} to which we often refer in the next section) the word ``vielbein'' is often used to denote the set of one forms $\{\tau_\mu, e_\mu{}^a\}$ rather than the vectors $\{n^{\mu}, e^{\mu}{}_a \}$.} We introduce the notation
\be \label{vielbein-def}
(\mathcal E_\m{}^A) \equiv  \left(  \t_\m, e_\m{}^a\right) , \qquad (\mathcal E^\m{}_A) \equiv  \left( n^\m, e^\m{}_a \right) ,
\ee
in terms of which the above relations read
\be
\mathcal E_\m{}^A \mathcal E^\m{}_B = \delta^A{}_B\, , \qquad \mathcal E_\m{}^A \mathcal E^\n{}_A = \delta_\m{}^\n \, .
\ee

The vielbein and its inverse enable one to convert spacetime indices into tangent space ones and vice-versa. We can in particular convert the indices of the Carroll-invariant internal metric $\zeta_{AB}$ and of the vector $n^A$ into spacetime indices, yielding a degenerate metric $g_{\mu\nu}$ in spacetime and the vector $n^\m$ already introduced in \eqref{vielbein-def},
\be
g_{\mu \nu} = \mathcal E_\mu{}^A \mathcal E_\nu{}^B \zeta_{AB} = e_\mu{}^a e_\nu{}^b \delta_{ab} \quad \text{and} \quad  n^\m = \cE^\m{}_A n^A \, ,
\ee
which fulfill by construction
\be
g_{\mu \nu} n^\nu = 0 \, .
\ee
It follows from this relation that $\tau_{\mu}$ is not obtained by lowering the index $\nu$ of $n^\nu$ with the metric $g_{\mu \nu}$ and this is the reason why we used a different letter. By contrast, we have $e_\mu{}^a\,\delta_{ab} = g_{\mu \nu}\,e^{\nu}{}_b $ and the use of the same letter here should not lead to confusion.  In fact, one has in all cases
\be
\zeta_{AB}\,\mathcal E_\mu{}^A = g_{\mu \nu}\,\mathcal E^{\nu}{}_B\,,
\ee
but this relation reduces to $0 = 0$ when $B= 0$.

Under local gauge transformations, the Carrollian connection \eqref{connection} transforms as $\delta A_\mu = \mathcal D_\mu \Gamma$, where $ \mathcal D_\mu \Gamma = \partial_\mu \Gamma + [A_\mu, \Gamma]$ is the covariant derivative of the gauge parameter 
\be
\Gamma = \x H + \x^a P_a + \la^a C_a + \frac{1}{2}\, \la^{ab} J_{ab} \,.
\ee
In components, this gives
\begin{subequations} \label{gauge-transf}
\begin{align}
\d e_\m{}^a &= \pr_\m \x^a + \omega_\m{}^{ab}\x_b - e_\m{}^b \la^a{}_b \,,  \\[4pt]
\d \t_\m &= \pr_\m \x + \omega_\m{}^a \x_a - e_\m{}^a \la_a \,, \label{var_tau} \\
\d \omega_\m{}^{ab} &= \pr_\m \la^{ab} + 2\,\omega_\m{}^{c[a} \la^{b]}{}_c - \frac{2\s}{\ell^2}\, e_\m{}^{[a} \x^{b]} \,,\\
\d \omega_\m{}^a &= \pr_\m \la^a + \omega_\m{}^{ab} \la_b - \omega_\m{}^b \la^a{}_b - \frac{\s}{\ell^2} \left(\t_\m \x^a - e_\m{}^a \x \right) ,
\end{align}
\end{subequations}
where $X_{[\mu}Y_{\nu]}=\frac{1}{2}\left(X_{\mu}Y_{\nu}-X_{\nu}Y_{\mu}\right)$. These formulas show that the determinant condition is preserved under transformations of the homogeneous Carroll group (boosts and spatial rotations, with $\xi = \xi_a = 0$) for which $\delta \mathcal E = 0$,  but in general not under inhomogeneous transformations (internal translations).

Given its importance, we write explicitly the transformation rule of the inverse vielbein to which \eqref{gauge-transf} reduces for  homogeneous Carroll transformations:
\be \label{gauge-transf-homogeneous0}
\d e_\m{}^a =  - e_\m{}^b \la^a{}_b \,,  \qquad
\d \t_\m = - e_\m{}^a \la_a \, . 
\ee
With the identification of $\{n^{\mu}, e^{\mu}{}_a \}$ as a basis  of the tangent space, we have also a linear action of the homogeneous Carroll group in the tangent space at each point, which follows  from
\eqref{gauge-transf-homogeneous0} and the duality of the frames, and reads
\be \label{gauge-transf-homogeneous1}
\d e^{\mu}{}_a =   e^{\mu}{}_b \la^b{}_a   + \la_a n^\mu\,,  \qquad
\d n^\m = 0 \,. 
\ee
It follows that, contrary to the one-form $\t_\mu$, the vector $n^\mu$ is invariant under local homogeneous Carroll transformations. The degenerate metric $g_{\mu \nu}$ is also invariant under these transformations in the tangent space since the internal metric \eqref{eq:internal_metric} is itself Carroll-invariant  ($\la_{ca} \equiv \delta_{cb} \la^b{}_a$ is antisymmetric).

One can write more compactly the transformation laws of the vielbein and the inverse vielbein as
\be
\d \mathcal E^{\mu}{}_A =   \mathcal E^{\mu}{}_B \la^B{}_A\, , \qquad \d \mathcal E_\m{}^A = - \mathcal E_\m{}^B \, \la^A{}_B
\ee
with
\be
(\la^A{}_B) = \begin{pmatrix} 0 & \la_b \\ 0 & \la^a{}_b  \end{pmatrix} .
\ee
Tensors in the tangent space transform with the matrix $\la^A{}_B$.

The metric $g_{\mu \nu}$ and the vector $n^\mu$ provide a complete set of Carroll invariants that can be constructed out of the vielbein. This is because $g_{\mu \nu}$ and $n^\mu$ determine the vielbein up to a Carroll transformation.  Indeed if $\{n'^{\mu}, e'^{\mu}{}_a \}$ is such that
\be
g_{\mu \nu} = e'_\mu{}^a e'_\nu{}^b \delta_{ab}\, , \qquad   n'^{\mu} = n^\mu \, ,
\ee
then $\{n'^{\mu}, e'^{\mu}{}_a \}$ differ from $\{n^{\mu}, e^{\mu}{}_a \}$ by a linear transformation preserving the Carrollian structure \eqref{eq:internal_metric}--\eqref{eq:internal_vector}, i.e., by a Carroll transformation. The invariants $g_{\mu \nu}$ and $n^\mu$ are redundant since one has $g_{\mu \nu} n^\mu = 0$.  A non-redundant set of variables is given by $(g_{\mu \nu}, \mathcal E)$: if one knows $g_{\mu \nu}$, the vector $n^\mu$ is determined up to normalization, which is fixed by $\mathcal E$. The variables $(g_{\mu \nu}, \mathcal E)$ are the basic variables of \cite{Henneaux:1979vn}, where $\mathcal E$ was denoted $\Omega$.
It follows from this observation that any function of the vielbein that is invariant under local transformations can be viewed as a function of $g_{\mu \nu}$ and $\mathcal E$. The field $\tau_\mu$ was also introduced in \cite{Henneaux:1979vn, Henneaux:2021yzg}, where it was denoted $\theta_\mu$.

\subsection{Extrinsic curvature and Carroll-compatible torsion-free connections}

Another important object introduced in \cite{Henneaux:1979vn}, which we will need below, is the second fundamental form or extrinsic curvature defined as\footnote{Note for comparison that the authors of ref.~\cite{Bergshoeff:2017btm} introduced an overall sign difference in the definition of the extrinsic curvature as compared to   \cite{Henneaux:1979vn}.}
\begin{equation}
K_{\mu\nu} \equiv -\frac{1}{2}\,\mathcal{L}_{n} g_{\mu\nu} = -\frac{1}{2}\left(n^{\rho}\partial_{\rho}g_{\mu\nu}+g_{\mu\rho}\partial_{\nu}n^{\rho}+g_{\nu\rho}\partial_{\mu}n^{\rho}\right) . \label{eq:extrinsic_curv}
\end{equation}
Contrary to what happens in the Riemannian case, it is now a spacetime tensor.   It has, however,  the same number of independent components as a spatial symmetric tensor since it is transverse,  $K_{\mu\nu} n^\nu = 0$.  Because of this property,  $K_{ab}$ defined through
\be
K_{ab} \equiv e^{\mu}{}_{a} e^{\nu}{}_{b} K_{\mu\nu}
\ee
contains the same information as $K_{\mu \nu}$ (since $K_{\mu \nu} = e_\m{}^a e_\nu{}^b K_{ab}$) and transforms as
\be
\delta K_{ab} = K_{cb} \la^c{}_a +K_{ac} \la^c{}_b \, 
\ee
under internal homogeneous Carroll transformations.

The identification of the components $\mathcal E_\mu{}^A$ of the Carrollian connection as a frame in the cotangent space enables one to define a covariant derivative $D_\mu$ for tangent tensors, by restricting the full Carroll-covariant derivative $\mathcal D_\mu$ to the homogeneous subgroup,
\be \label{hom-cov-der}
D_\mu T = \partial_\mu T + [\omega_{\mu}, T]\, , \qquad \omega_{\mu} = \omega_\mu{}^a C_{a}+\frac{1}{2}\,\omega_\mu{}^{ab} J_{ab} = \frac{1}{2}\,\omega_\mu{}^{AB} J_{AB}
\ee
($\omega_\mu{}^a \equiv \omega_\mu{}^{0a} = - \omega_\mu{}^{a0}$). For instance, for a vector $v = v^A P_A = v^0 H + v^a P_a$, 
\be
D_\mu v^0 = \partial_\mu v^0 + \omega_{\mu\,a} v^a \,, \qquad D_\mu v^a = \partial_\mu v^a + \omega_\mu{}^{a}{}_b v^b
\ee
and similarly for a covector $\psi_A$
\be
D_\mu \psi_0 = \partial_\mu \psi_0 \,,  \qquad D_\mu \psi_a = \partial_\mu \psi_a -  \omega_{\mu\,a} \psi_0 + \omega_{\mu\,a}{}^{b} \psi_b \,.
\ee
This covariant derivative automatically preserves the metric and the normal vector $n^A$
\be
D_\mu \zeta_{AB} = 0 \, , \qquad D_\mu n^A = 0 \, .
\ee
In analogy with the terminology used in Riemannian geometry, one might call $\omega_\mu{}^{AB}$ introduced in eq.~\eqref{hom-cov-der} the ``spin connection''.  

Parallel transport and covariant derivatives are concepts that can be formulated in any tangent basis, so one can translate the covariant derivatives defined in the local frame $\{\mathcal E^\mu{}_A \frac{\partial}{\partial x^\mu} \} $ to the coordinate frame $\{ \frac{\partial}{\partial x^\mu} \} $. The connection in the coordinate basis is denoted $\Gamma^{\mu}{}_{\rho \sigma}$.  The spin connection $\omega_\mu{}^{AB}$ and $\Gamma^{\mu}{}_{\rho \sigma}$ are related through the standard change-of-frame formulas for a connection, namely
\begin{subequations} \label{Gamma-Omega}
\begin{align}
\partial_\rho \mathcal E_\mu{}^0 + \omega_\rho{}^{a} \mathcal E_\mu{}_a - \Gamma^{\sigma}{}_{\rho \mu} \mathcal E_{\sigma}{}^0 & = 0 \, , \\[5pt]
\partial_\rho \mathcal E_\mu{}^a + \omega_\rho{}^{ab} \mathcal E_\mu{}_b - \Gamma^{\sigma}{}_{\rho \mu} \mathcal E_{\sigma}{}^a & = 0 \, .
\end{align}
\end{subequations}
These can be compactly written
\be
D_\rho \mathcal E_\mu{}^A = 0 \, ,
\ee
where $D_\rho$ acts on all (internal and spacetime) indices. One way to think about these formulas is that they express that the covariant derivative of the Kronecker tensor is zero (as it should!), in particular if the computation is carried in a mixed basis (one index in the local frame, one index in the coordinate frame).  One sometimes calls these equations ``the vielbein postulate''. 

Torsion-free connections, for which $D_{[\mu} \chi_{\nu]} =  \partial_{[\mu} \chi_{\nu]}$ for any one-form $\chi_\mu$,  play an important role in Riemannian geometry and are defined in an analogous manner in Carrollian geometry. In a coordinate frame, in both Riemannian and Carrollian geometry, the absence of torsion is equivalent to the symmetry of $\Gamma^{\sigma}{}_{\rho \mu}$ in its lower indices $\rho, \mu$,
\be \label{torsion-free}
\Gamma^{\sigma}{}_{\rho \mu}  = \Gamma^{\sigma}{}_{\mu \rho } \, .
\ee
For Carrollian geometry, in the local frames $\{\cE^\mu{}_A\}$ this condition is itself equivalent in view of \eqref{Gamma-Omega} to the vanishing of the torsion tensors\footnote{In order to connect our notation with that of \cite{Bergshoeff:2017btm}, one has to change the sign of $\omega_\mu{}^a$, $\omega_\mu{}^{ab}$ and of the tensors $R_{\mu\nu}{}^{a}$ and $R_{\mu\nu}{}^{ab}$ that we will introduce shortly in \eqref{curvatures}. This is a consequence of a sign difference in the generators $C_a$ and $J_{ab}$ of the Carroll algebra \eqref{Carroll-algebra}.} 
\begin{subequations} \label{torsions}
\begin{align}
T_{\mu\nu} &= 2 \left( \partial_{[\mu}\tau_{\nu]} +  \omega_{[\mu}{}^{a}e_{\nu]a} \right) , \label{eq:Torsion1}\\[5pt]
T_{\mu\nu}{}^{a} &= 2 \left( \partial_{[\mu}e_{\nu]}{}^{a} +  \omega_{[\mu}{}^{ab}e_{\nu]b} \right) , \label{eq:Torsion2}
\end{align}
\end{subequations}
expressions which can be succinctly written as
\be
T_{\mu\nu}{}^{A} = 2 \left( \partial_{[\mu}\cE_{\nu]}{}^{A} +  \omega_{[\mu}{}^{AB}\cE_{\nu]}{}^C \zeta_{BC} \right) . \label{eq:Torsion3}
\ee

It is a well-known result of Riemannian geometry that there is a unique torsion-free, metric-compatible connection for any Riemannian metric, called the ``Levi-Civita connection''. This is not so in Carrollian geometry. One has instead \cite{Henneaux:1979vn, Vogel1965, Jankiewicz, Dautcourt1967}:
\begin{itemize}
\item A necessary and sufficient condition for the existence of a torsion-free connection that preserves the Carrollian structure ($D_\rho g_{\mu \nu} = 0$, $D_\rho n^\mu = 0$) is that the extrinsic curvature vanishes,
\be
K_{\mu \nu} = 0 \, .
\ee
\item When this condition is satisfied, the connection is not unique but determined up to the addition of $n^\rho S_{\mu \nu}$ where $ S_{\mu \nu}$ is an arbitrary symmetric and transverse tensor,
\be
\Gamma^{\rho}{}_{\mu \nu}  \rightarrow \Gamma^{\rho}{}_{\mu \nu} + n^\rho S_{\mu \nu}\, , \qquad S_{\mu \nu} = S_{\nu \mu} \, , \qquad S_{\mu \nu} n^\mu = 0 \, .
\ee
\end{itemize}
The proof is direct and is most easily carried out in local coordinates where $n^{\mu} = (1, 0, \cdots, 0)$. [In such a coordinate system, $g_{tt} = g_{ti} = 0$ and $\Gamma^{k}{}_{ij} = \gamma^{k}{}_{ij}$ are the
spatial Christoffel symbols, $\Gamma^{j}{}_{ti} = 0$ ($\Leftrightarrow K_{ij} = 0$) while the $\Gamma^{t}{}_{ij}$ are arbitrary.] Another way to see that the extrinsic curvature must vanish is to consider the components 
\be
T_{ABC} \equiv \mathcal E^\mu{}_A \mathcal E^\nu{}_B T_{\mu\nu}{}^{D}\zeta_{CD}
\ee
of the torsion tensor and to observe that the connection drops from $T_{0(bc)}$, which contains only the vielbein and its derivative and is in fact proportional to $K_{\mu \nu}  \mathcal E^\mu{}_b \mathcal E^\nu{}_c$, so that the constraint $T_{0(bc)}= 0$ implied by the torsion-free condition \eqref{torsion-free}, enforces $K_{\mu \nu} = 0$. 

\subsection{Torsion and curvature}

The curvature of the Carrollian connection \eqref{connection} can be decomposed as 
\be \label{curvature}
\cF_{\mu\nu} = \partial_\mu A_\nu - \partial_\nu A_\mu + [A_\mu, A_\nu] = T_{\mu\nu}^{\phantom{a}}\, H + T_{\mu\nu}{}^a P_a + F_{\mu\nu}{}^a C_a + \frac{1}{2} F_{\mu\nu}{}^{ab} J_{ab} \, ,
\ee
where the ``torsions'' $T_{\m\n}$ and $T_{\m\n}{}^a$ are defined in eq.~\eqref{torsions} and where the ``curvatures'' read
\begin{subequations} \label{curvatures}
\begin{align}
F_{\mu\nu}{}^{a} &= 2 \left( \partial_{[\mu}\omega_{\nu]}{}^{a} + \omega_{[\mu}{}^{ab}\omega_{\nu]b} - \frac{\sigma}{\ell^2}\, \tau_{[\mu}e_{\nu]}{}^a \right) \equiv
R_{\mu\nu}{}^{a} - \frac{2\sigma}{\ell^2}\, \tau_{[\mu}e_{\nu]}{}^a \,,\label{eq:curvature1} \\
F_{\mu\nu}{}^{ab} &= 2 \left( \partial_{[\mu}\omega_{\nu]}{}^{ab} + \omega_{[\mu}{}^{ac}\omega_{\nu]\,c}{}^{b} -\frac{\sigma}{\ell^2}\, e_{[\mu}{}^a e_{\nu]}{}^b \right) \equiv R_{\mu\nu}{}^{ab} -\frac{2\sigma}{\ell^2}\, e_{[\mu}{}^a e_{\nu]}{}^b \,\label{eq:curvature2}.
\end{align}
\end{subequations}
Here, we also introduced the curvatures of the Carroll algebra, denoted by $R_{\mu\nu}{}^{a}$ and $R_{\mu\nu}{}^{ab}$, that one recovers for $\sigma = 0$.  They read explicitly
\be
R_{\mu\nu}{}^{AB} = 2 \left( \partial_{[\mu}\omega_{\nu]}{}^{AB} + \omega_{[\mu}{}^{AC}\omega_{\nu]}{}^{DB}\zeta_{CD}\right) .\label{eq:curvature3}
\ee
Note that because the invariant metric $\zeta_{DB}$ is degenerate, there is no term quadratic in $\omega_{\mu}{}^{a}$.

The full curvature in eq.~\eqref{curvature} transforms covariantly under Yang-Mills-type gauge transformations, which yields in terms of its components
\begin{subequations} \label{gauge-transf-curvatures}
\begin{align}
\d T_{\m\n}{}^a &=  F_{\mu \nu}{}^{ab}\x_b - T_{\m\n}{}^b \la^a{}_b \,,  \\[5pt]
\d T_{\m\n} &=  F_{\mu\nu}^{a}  \x_a - T_{\m\n}{}^a \la_a \,,  \\
\d F_{\m \n}{}^{ab} &=  2\,F_{\m \n}{}^{c[a} \la^{b]}{}_c - \frac{2\s}{\ell^2}\, T_{\m\n}{}^{[a} \x^{b]} \,,\\
\d F_{\m\n}{}^a &=  F_{\m\n}{}^{ab} \la_b -F_{\m\n}{}^b \la^a{}_b - \frac{\s}{\ell^2} \left(T_{\m\n} \x^a - T_{\m\n}{}^a \x \right) .
\end{align}
\end{subequations}
While torsion and curvature mix under Carrollian translations, they transform separately under the homogeneous Carroll subgroup, i.e.\ for $\x_b = 0$ and $\x = 0$: 
\begin{subequations} \label{gauge-transf-curvatures-Bis}
\begin{alignat}
\d T_{\m\n} {}^a &=  - T_{\m\n}{}^b \la^a{}_b \,, \qquad 
& \d T_{\m\n} &=   - T_{\m\n}{}^a \la_a \,, \\
\d F_{\m \n}{}^{ab} &=  2\,F_{\m \n}{}^{c[a} \la^{b]}{}_c  \,, 
\qquad
& \d F_{\m\n}{}^a &=  F_{\m\n}{}^{ab} \la_b -F_{\m\n}{}^b \la^a{}_b  \,  ,
\end{alignat}
\end{subequations}
implying 
\begin{equation} \label{gauge-transf-curvatures-Ter}
\d R_{\m \n}{}^{ab} =  2\,R_{\m \n}{}^{c[a} \la^{b]}{}_c  \,, \qquad
\d R_{\m\n}{}^a = R_{\m\n}{}^{ab} \la_b - R_{\m\n}{}^b \la^a{}_b  \, .
\end{equation}

Setting
\be 
(T_{\m\n} {}^A) = (T_{\m\n}, T_{\m\n} {}^a)\, , \qquad R_{\m \n}{}^{0a}= R_{\m\n}{}^a \, , \qquad 
\ee
these relations can be written as 
\be
\d T_{\m\n} {}^A = -  \la^A{}_B T_{\m\n} {}^B\, , \qquad \d  R_{\m \n}{}^{AB} = -  \la^A{}_C R_{\m \n}{}^{CB}  -  \la^B{}_C R_{\m \n}{}^{AC} \, .
\ee
It follows in particular from these relations that 
\be
 \mathcal R = \mathcal E^{\mu}{}_A  \mathcal E^{\nu}{}_B R_{\m \n}{}^{AB}
\ee
is invariant both under local Carroll homogeneous transformations in the tangent space and coordinate transformations.  It is called the scalar curvature.\footnote{We use the notation $\mathcal R$ not to confuse this object with the spatial scalar curvature appearing in the Hamiltonian formulation.}  In components,
\be
\mathcal R = 2\, n^{\mu}e^{\nu}{}_{a} R{}_{\mu\nu}{}^{a} + e^{\mu}{}_{a} e^{\nu}{}_{b} R_{\mu\nu}{}^{ab} \, .
\ee

\section{Gauging of the Carroll algebra - Dynamics}\label{sec:gauging2}

In this section, we first review the action proposed in \cite{Bergshoeff:2017btm} and we highlight that its equations of motion select a torsion-free Carrollian connection. As we have seen, this connection is defined up to an arbitrary symmetric tensor and we show that the latter can be identified with the momentum conjugate to the spatial metric. Taking advantage of this observation, we then prove that the action obtained from the gauging of the Carroll algebra in \cite{Bergshoeff:2017btm} is equivalent to the action of magnetic Carrollian gravity of \cite{Henneaux:2021yzg}.

\subsection{Action from gauging}

To fix the dynamics one can set up an action principle requiring that the action be invariant only under local homogeneous Carroll transformations. This condition preserves the non-degeneracy assumption. One also requires the action to be invariant under spacetime diffeomorphisms (which preserve $\mathcal E \not=0$ too).
One obvious candidate, which is the analog of the Einstein-Hilbert action ---~or rather, its first-order Einstein-Cartan formulation in which vielbein and spin connection are treated as independent variables~---, is 
\begin{equation} 
\begin{split}
I_{\hbox{\tiny{Car}}}[\mathcal E_\mu{}^A, \omega_\mu{}^{AB}] &= \frac{1}{16\pi G_M}\int dt\,d^{D}x\, \mathcal E \left(\mathcal R - 2 \Lambda\right) \\
& = \frac{1}{16\pi G_M}\int dt\,d^{D}x\, \mathcal E \left(2\,n^{\mu}e^{\nu}{}_{a} R{}_{\mu\nu}{}^{a} + e^{\mu}{}_{a} e^{\nu}{}_{b} R_{\mu\nu}{}^{ab} - 2 \Lambda \right) .\label{eq:first-order-action}
\end{split}
\end{equation}
This is the action proposed in \cite{Bergshoeff:2017btm} starting from a $c \to 0$ limit of the relativistic Einstein-Cartan action, where $G_M$ is a constant resulting from a rescaling of Newton's constant.\footnote{When $D+1=4$ this action can be written in a MacDowell-Mansouri form: $I_{\hbox{\tiny{Car}}}=-\frac{\sigma\ell^2}{16\pi G_M}\int \left\langle \cF\wedge \cF\right\rangle$ with $\left\langle C_a J_{bc}\right\rangle=\epsilon_{abc}$.} See also \cite{Figueroa-OFarrill:2022mcy} for a classification of all terms built out of the connections in eq.~\eqref{connection} and the curvatures in eq.~\eqref{curvature} that are invariant under local homogeneous Carroll transformations.

The equations of motion that one derives from the action \eqref{eq:first-order-action} impose the vanishing of the torsion \eqref{eq:Torsion3} (by extremizing the action with respect to the spin connection), 
together with a Carrollian analog of Einstein's equations (by extremizing the action with respect to the vielbein). 
Thus, the action \eqref{eq:first-order-action} forces the connection to be torsion-free in addition to preserving the Carrollian structure.  From what we have recalled above, this implies that the extrinsic curvature vanishes and that the connection is not uniquely determined from the vielbein but involves an arbitrary, transverse symmetric tensor.

In order to see how this arises in detail, we follow closely, in this paragraph and the next,  the paper \cite{Bergshoeff:2017btm}, which provides important insight into the dynamical aspects of the theory with action \eqref{eq:first-order-action}.  From the torsion equations, one can solve for all the connection components, except for the symmetrized boost component $\omega_{(ab)}{}^0 \equiv \mathcal E^\mu{}_{(a} \delta_{b)c}\omega_\mu{}^{c0}$ of the connection \eqref{hom-cov-der}, which remains arbitrary and in terms of which the other components can be expressed,
\begin{subequations} \label{sol-torsions}
\begin{align}
\omega_{\mu}{}^{a} & =-\,\tau_{\mu}n^{\nu}e^{\rho\, a}\partial_{[\nu}\tau_{\rho]}-e^{\nu \,a}\partial_{[\mu}\tau_{\nu]}+S^{ab}e_{\mu\,b}\,,\\[5pt]
\omega_{\mu}{}^{ab} & =2\,e^{\rho[a}\partial_{[\mu}e_{\rho]}{}^{b]}-e_{\mu\,c}e^{\rho\, a}e^{\nu\, b}\partial_{[\rho}e_{\nu]}{}^{c}\,.
\end{align}
\end{subequations}
Here, $S^{ab} = S^{ba}$ is an arbitrary symmetric tensor, which can be identified with $-\omega^{(ab)0}$ as one can verify immediately from  \eqref{sol-torsions}. One can view $S^{ab}$ as the spatial components of a symmetric tensor $S^{AB}$ in the vielbein basis, the components $S^{0A}$ of which are arbitrary but do not occur in the above expressions. By lowering the indices, one gets a symmetric, transverse, twice covariant tensor, which captures the ambiguity in the Carroll-compatible, torsion-free connections described above.  
The variation of the action with respect to $\omega^{(ab)0}$ yields an equation equivalent to $T_{0(ab)} = 0$, which cannot be solved for $\omega_{(ab)}{}^0$ but sets instead the extrinsic curvature to zero.

The next step is to eliminate the dependent spin connection components 
using \eqref{sol-torsions} while keeping the independent ones $S^{ab}$, to get \cite{Bergshoeff:2017btm}
\be \label{eq:action2}
\begin{split}
I_{\hbox{\tiny{Car}}}[\mathcal E_\mu{}^A, S^{ab}] =\frac{1}{16\pi G_M}\int dt\,d^{D}x\, \mathcal E\, \Big(& 2\,n^{\mu}e^{\nu}{}_{a}\!\left. R_{\mu\nu}{}^{a}\right|_{S^{ab}=0}+e^{\mu}{}_{a} e^{\nu}{}_{b} R_{\mu\nu}{}^{ab} - 2\Lambda \\
& - 2\left(S^{ab}-\delta^{ab} S \right) K_{ab} \Big) \, ,
\end{split}
\ee
where $S \equiv \delta_{ab}S^{ab}$. In this form of the action, it is manifest that the field $S^{ab}$ acts as a Lagrange multiplier enforcing the condition
\be \label{K=0}
K_{ab}=0
\ee
that the extrinsic curvature of the metric $g_{\m\n}$ should vanish.  

\subsection{Time gauge}

We have now everything at hand to establish the equivalence of the action \eqref{eq:action2} with the action \eqref{eq:magnetic_action} describing the magnetic limit of Einstein's theory.  The most expedient way to do so is to (i) go to the ``time gauge'', i.e., use the freedom in the Carrollian boosts to set 
\be
\tau_i = 0 \, ,
\ee
where we split Greek indices as $\mu=\{t,i\}$ (this is permissible since the theory is invariant under local homogeneous Carroll transformations); and (ii) introduce the lapse $N$ and the shift $N^i$ familiar in the ADM ($D+1$)-decomposition of general relativity, which express the vector $\frac{\partial} {\partial t}$ in the frame $\{n^\mu \frac{\partial}{\partial x^\mu} , \frac{\partial}{\partial x^i} \}$.\footnote{We assume that the hypersurfaces $t=$ constant are transverse to the integral curves of $n^\mu$, so that the metric $g_{ij}$ induced on these hypersurfaces is non-degenerate and of Euclidean signature.} We then find that the vielbein and the inverse vielbein are parametrized as
\be
n^{\mu}= \left(\frac{1}{\lapse},-\frac{N^{i}}{\lapse}\right) , \qquad 
e^{\mu}{}_{a} =\left(0,\frame^i{}_{a}\right) ,
\ee
and
\be \label{Carroll-frame}
\tau_{\mu}= \left(\lapse,0\right) , \qquad e_{\mu}{}^{a}=\left(\frame_{i}{}^{a}N^{i},\frame_{i}{}^{a}\right) ,
\ee
respectively, where $\frame^i{}_{a}$ is a spatial local frame with inverse $\frame_{i}{}^{a}$ (``$D$-bein''),
\be \frame^i{}_{a} \frame_{i}{}^{b} = \delta_a{}^b \, , \qquad  
\frame^i{}_{a} \frame_{j}{}^{a} = \delta^i{}_j \, .
\ee
These expressions take the same form as in the customary time gauge in general relativity \cite{Dirac1962, Nelson:1978ex, Henneaux:1978wlm}. The residual local Carroll freedom is exhausted by the local rotations acting on the spatial $D$-bein.

In terms of this parametrization, useful formulas are
\be \label{metric_components}
g_{ij} = \frame_{i}{}^{a}\frame_{j}{}^{b}\delta_{ab} \,,\quad g_{ti} = g_{ij}N^{j} \,,\quad g_{tt} = g_{ij} N^{i} N^{j} \, ,
\ee
for the metric,
\be
\mathcal E = N \frame = N \sqrt{g} \, \qquad \hbox{($g=$ determinant of spatial metric)}
\ee
for the determinant of the vielbein,
\be \label{eq:extrinsic_curv_constraint}
K_{ij} = -\frac{1}{2\lapse}\left(\dot{g}_{ij}-\nabla_{\!i}N_{j}-\nabla_{\!j}N_{i}\right), \quad K_{ti} = K_{ij} N^j \,,\quad K_{tt} = K_{ij} N^i N^j \,,
\ee
for the extrinsic curvature (where $\nabla$ denotes the Levi-Civita connection for the spatial metric $g_{ij}$) and, using also the time-gauge,
\begin{subequations}\label{spin-conn-expl}
\begin{align}
\omega_t{}^a &= \frame^i{}^a \partial_{i} \lapse+ N^i S^{a}_{\;\,b} \frame_i{}^b\,,
\\
\omega_i{}^a &= S^{a}_{\;\,b} \frame_i{}^b \,,
\\
\omega_t{}^{ab} &= \frame^j{}^{[a}\dot{\frame}_{j}{}^{b]} -  \frame^j{}^{[a}  \partial_{j} ( \frame_{i}{}^{b]} N^i) - N^i \frame^j{}^a \frame^k{}^b \frame_i{}_c \partial_{[j} \frame_{k]}{}^c,
\\
\omega_i{}^{ab} &= 2\, \frame^j{}^{[a} \partial_{[i} \frame_{j]}{}^{b]} - \frame^j{}^a \frame^k{}^b \frame_i{}_c \partial_{[j} \frame_{k]}{}^c,
\end{align}
\end{subequations}
for the spin connection.

\subsection{Recovering the magnetic action}

Substituting eqs.~\eqref{spin-conn-expl} in the expression \eqref{eq:curvature3} for the curvature gives  
\be
2\,n^\m e^\n{}_a \left.R_{\mu\nu}{}^{a}\right|_{S^{ab}=0}=-\,\partial_{i}\left(2\,\frame\,\frame^i{}_a \frame^j{}^a\partial_{j} N\right) , \qquad e^\m{}_a e^\n{}_b R_{\mu\nu}{}^{ab} = R\,,\label{eq:matching_of_curvatures}
\ee
where $R$ is the Ricci scalar constructed with the spatial metric $g_{ij}$. Dropping a total derivative, the action \eqref{eq:action2} becomes then
\be
I_{\hbox{\tiny{Car}}}[\frame_i{}^a, N, N^i, S^{ij}]  = \frac{1}{16\pi G_M} \int dt\,d^{D}x\, \sqrt{g}\, N  \left[  R  -2  \La  -2 \left(S^{ij}-\delta^{ij} S \right) K_{ij}
 \right], \label{eq:ActionENNS1}
\ee
where we have also made the change of dynamical variables $S^{ab} \rightarrow S^{ij} = \frame^i{}_a \frame^j{}_b S^{ab}$.  The action \eqref{eq:ActionENNS1} is manifestly first order in time derivatives (through $K_{ij}$) and second order in spatial derivatives (through $R$). 

It is also manifest that the action involves the $D$-bein only through the spatial metric.  This is of course a manifestation of local rotation invariance.   So, one really has $I_{\hbox{\tiny{Car}}}[\frame_i{}^a, N, N^i, S^{ij}] \equiv I_{\hbox{\tiny{Car}}}[g_{ij}, N, N^i, S^{ij}]$ with exactly the same integral
\be
I_{\hbox{\tiny{Car}}}[g_{ij}, N, N^i, S^{ij}]  = \frac{1}{16\pi G_M} \int dt\,d^{D}x\, \sqrt{g}\, N  \left[  R  -2  \La  -2\left(S^{ij}-\delta^{ij} S \right)K_{ij}
 \right]. \label{eq:ActionENNS2}
\ee
Extremization with respect to variations of the spatial metric automatically extremizes with respect to $D$-bein variations.

We are now at a stone throw from proving equivalence.   This is just achieved by making one more change of variables, namely,
\be \label{eq:S_to_pi}
\pi^{ij}=\frac{\sqrt{g}}{16\pi G_M}\,\left(S^{ij}-h^{ij}S\right)  \quad \Leftrightarrow \quad S^{ij}=\frac{16\pi G_M}{\sqrt{g}}\left(\pi^{ij}-\frac{1}{D-1}\,h^{ij}g_{kl}\pi^{kl}\right) ,
\ee
where $h^{ij}$ is the inverse to the spatial metric $g_{ij}$. Using the explicit expression of  the extrinsic curvature in terms of $\dot{g}_{ij}$, the lapse, and the shift in eq.~\eqref{eq:extrinsic_curv_constraint}, and making an integration by parts gives then immediately 
\be
I_{\hbox{\tiny{Car}}}[g_{ij}, N, N^i, S^{ij}] =  \int dt d^D x \left(\pi^{ij} \dot{g}_{ij} + 2 N_i \nabla_j \pi^{ij} - N \mathcal H_M \right) , \label{eq:magnetic-ADM}
\ee
and therefore the searched-for equality
\be
I_{\hbox{\tiny{Car}}} = I_M
\ee
of the action \eqref{eq:first-order-action} of \cite{Bergshoeff:2017btm} (after the successive transformations explained above) and the magnetic action \eqref{eq:magnetic_action} of \cite{Henneaux:1979vn}.

\section{Magnetic limit of the Einstein-Cartan action}\label{sec:limit}

In the previous section we related directly the first-order action \eqref{eq:first-order-action} (or its equivalent reformulation \eqref{eq:action2}) of \cite{Bergshoeff:2017btm} to the action of magnetic Carrollian gravity in ADM form of \cite{Henneaux:2021yzg}. In both cases, we thus directly considered the result of a ultra-relativistic limit of general relativity. In this section, we instead first recall the relation between the first-order and ADM formulations of general relativity following \cite{Pilati:1977ht, Nelson:1978ex, Castellani:1981ue} and then we track the effect of the limit at each stage of the computations. The key point we wish to highlight is that in the relativistic case one can choose either to fully eliminate the spin connection via the torsion constraints to recover the second-order formulation of general relativity or to keep suitable components of the spin connection as independent fields. These turn out to be proportional to the conjugate momenta to the spatial metric as in \eqref{eq:S_to_pi} and one eventually recovers the ADM formulation of general relativity. After the limit $c \to 0$ is taken, the first option is instead not available anymore and, as we discussed in the previous section, one is forced to keep the field $S^{ab}$ in the action.

\subsection{Setting the stage for the Carrollian limit}

We start from the first-order formulation of general relativity, i.e.\ from the Einstein-Cartan action
\be \label{Einstein-Cartan}
I = \frac{c^3}{16 \p G_N} \, \int  dt \, d^D x \ E \left( E^\m{}_A\,E^\n{}_B\, \cR_{\m\n}{}^{AB} - 2\,\La \right) ,
\ee
where
\be \label{Lorentz-curvature}
\cR_{\m\n}{}^{AB} = 2\,\pr_{[\m}\,\O_{\n]}{}^{AB} + 2\,\O_{[\m}{}^{AC}\,\O_{\n]}{}^{DB}\,\h_{CD} \, ,
\ee
with $\h_{CD} = \text{diag}(-1,+1,\ldots,+1)$, while $c$ denotes the speed of light and $G_N$ is Newton's constant. Moreover, tangent space indices take the values $A,B=0,1,\dots,D$, while $E_\m{}^A$ is the coframe, $\Omega_\m{}^{AB}$ is the spin connection and $E=\text{det}\left(E_\m{}^A \right)$. To define the Carrollian limit we introduce a dimensionless parameter $\e$ via $c = \e\,\hat c$, so that the limit corresponds to sending $\e \to 0$. For simplicity, in the following we shall also set $\hat c = 1$. 
We then consider the same scaling in $\epsilon$ for the components of the coframe and of the spin connection as in \cite{Bergshoeff:2017btm}:
\be \label{rescalings}
E_\m{}^A
= \left( \e\, \t_\m, e_\m{}^a\right), \qquad
\O_\m{}^{AB} 
= \left( \e\,\omega_\m{}^a, \omega_\m{}^{ab} \right) ,
\ee
where we make explicit the link with the one-forms that we used in the previous sections. Taking the limit $\epsilon \to 0$ while rescaling Newton's constant as
\be \label{rescale_Newton}
G_N = \e^{4}\,G_M
\ee
leads to the action \eqref{eq:first-order-action}.

To link the first-order action \eqref{Einstein-Cartan} to the ADM formulation of general relativity, it is convenient to introduce the following quantities starting from the components of the rescaled coframe $E_\m{}^A$ of eq.~\eqref{rescalings} and its inverse $E^\m{}_A$:
\be \label{eq:basis}
e_i{}^A \equiv E_i{}^A
\, , \qquad 
\text{n}_A \equiv -\e\,\lapse\,E^t{}_A \,,
\ee
where $N$ is the usual lapse function, while the covector $\text{n}_A$ at this stage should not be confused with the vector $n^A$ defined in eq.~\eqref{eq:internal_vector}. The variables $e_i{}^A$ and $\text{n}_A$ satisfy the relations
\be
\h^{AB}\,\text{n}_A\,\text{n}_B = -1 \,, \qquad e_i{}^A\,\text{n}_A = 0 \,,
\ee
where capital Latin indices are raised and lowered with the Minkowski metric $\h_{AB}$.
All components of the vielbein and its inverse are determined in terms of $e_i{}^A$, $\text{n}_A$ and the functions $N$ and $N^i$:
\be \label{vielbein}
E^\mu{}_A = (-\e^{-1}\,N^{-1}\,\text{n}_A ,\ e^i{}_A + \e^{-1}\,N^{-1}\,N^i\,\text{n}_A) \,,
\quad
E_\mu{}^A = (\e\,N\,\text{n}^A + e_i{}^A N^i ,\ e_i{}^A)\,,
\ee
where we introduced the quantity $e^i{}_A$ verifying
\be \label{completeness}
e^i{}_A\,e_j{}^A = \d^i{}_j \, , \qquad
e^i{}_A\,e_i{}^B = \d_A{}^B + \text{n}_A \text{n}^B \,.
\ee
The latter can also be defined as $e^i{}_A=h^{ij}e_j{}^B\,\h_{AB}$, where $h^{ij}$ is the inverse of the spatial metric
\be \label{spatial_metric}
g_{ij} = e_i{}^A\,e_j{}^B\,\h_{AB} \,.
\ee
The parametrization \eqref{vielbein} of the vielbein and its inverse implies the usual ADM decomposition of the metric:
\be \label{ADM-metric}
g_{\m\n} = \begin{pmatrix} N^i N_i - \e^2\,\lapse^2 & N_i \\ N_i & g_{ij} \end{pmatrix} , \qquad 
g^{\m\n} = \begin{pmatrix} - \frac{1}{\e^2\,\lapse^2} & \frac{N^i}{\e^2\,\lapse^2} \\ \frac{N^i}{\e^2\,\lapse^2} & h^{ij} - \frac{N^i N^j}{\e^2\,\lapse^2} \end{pmatrix} ,
\ee
where the spatial indices $i,j = 1, \ldots, D$ are raised and lowered using the $D$-dimensional spatial metric $g_{ij}$ and its inverse $h^{ij}$. From eq.~\eqref{ADM-metric}, it is clear that the metric $g_{\m\n}$ becomes degenerate in the limit $\e \to 0$.

All in all, together with the identity $E = \e\,\lapse \sqrt{g}$ (where we recall that $\sqrt{g}$ is the determinant of the spatial metric $g_{ij}$), this parametrization implies the following rewriting of the Einstein-Cartan action,
\be \label{NL:EHC}
\begin{split}
I = \e^3\! \int & \frac{dt d^D x\sqrt{g}}{16 \p G_N} \, \Big[2\,\text{n}_{[B}\,e^i{}_{A]}\,\cR_{ti}{}^{AB} + \e\,\lapse \left( e^i{}_{[A}\,e^j{}_{B]}\cR_{ij}{}^{AB} - 2\, \La \right) + 2\,N^{[i}\,\text{n}_{[A}\,e^{j]}{}_{B]} \cR_{ij}{}^{AB} \Big]\, .
\end{split}
\ee

\subsection{Torsion constraint}

Varying the action \eqref{NL:EHC} with respect to the full spin-connection $\Omega_\m{}^{AB}$ imposes the vanishing of the torsion
\be
\cT_{\m\n}{}^A = 2\, \pr_{[\m} E_{\n]}{}^A + 2\, \O_{[\m}{}^{AB} E_{\n]}{}^C \h_{BC} \, .
\ee
We now wish to consider the variation of the same action with respect to the different components of the spin connection, in order to identify which components of the torsion are set to zero by each of these variations. In particular, we distinguish
\begin{subequations}
\begin{alignat}{5}
\O_t{}^{ij} & = \O_t{}^{AB}\,e^i{}_A\,e^j{}_B \,, \qquad & \O_t{}^i{}_\perp & = \O_t{}^{AB}\,e^i{}_A\,\text{n}_B \,, \\[2pt]
\O_i{}^{jk} & = \O_i{}^{AB}\,e^j{}_A\,e^k{}_B \,, \qquad & \O_{ij\perp} & = \O_i{}^{AB}\,e_j{}_A\,\text{n}_B \,.
\end{alignat}
\end{subequations}
We can also project the components of the torsion on the basis of the tangent space introduced in eq.~\eqref{eq:basis} and obtain the following relevant components:
\begin{subequations}\label{components-torsion}
\begin{align}
\cT_{ij}{}_\perp & \equiv \cT_{ij}{}^A\,\text{n}_A = 2\,\pr_{[i}\,e_{j]}{}^A\,\text{n}_A - 2\,\O_{[ij]\perp} \,,\\[5pt]
\cT_{ij}{}^k & \equiv \cT_{ij}{}^A\,e^k{}_A = 2\,\pr_{[i}\,e_{j]}{}^A\,e^k{}_A - 2\,\O_{[ij]}{}^k \,,\\[5pt]
\cT_{ti}{}_\perp & \equiv \cT_{ti}{}^A\,\text{n}_A = \dot e_i{}^A\,\text{n}_A - N^j\,\pr_i\,e_j{}^A\,\text{n}_A - \O_{ti\perp} + N^j\,\O_{ij\perp} + \e\,\pr_i \lapse \,,\\[5pt]
\cT_{t[ij]} & \equiv \cT_{t[i}{}^A\,e_{j]}{}_A = \dot e_{[i}{}^A \, e_{j]}{}_A - \e\,\lapse\,\pr_{[i}\,e_{j]}{}^A\,\text{n}_A - e_{[i}{}^A\,\pr_{j]}\,N^k\,e_k{}_A \nonumber\\
&\qquad \qquad \qquad - N^k\,e_{[i}{}^A\,\pr_{j]}\,e_k{}_A - \O_{tij} + \e\,\lapse\,\O_{[ij]\perp} + N^k\,\O_{[ij]k} \,.
\end{align}
\end{subequations}
As we shall see in detail in the following, the components of the torsion that we wrote explicitly are set to zero by the equations of motion that follow from the variation of the action \eqref{NL:EHC} with respect to all components of the spin connection, except $\Omega_{(ij)\perp}$. This component plays a role similar to that of the tensor $S^{ab}$ in the previous section and it is related to the conjugate momenta of the spatial metric in the Hamiltonian formulation. The corresponding ``missing'' torsion constraint $\cT_{t(ij)}$ corresponds to the equation of motion allowing one to solve for $\pi^{ij}$ in terms of the time derivative of the spatial metric. 

In particular, varying with respect to $\O_t{}^{ij}$ and $\O_t{}^i{}_\perp$ (which are both Lagrange multipliers) one gets, respectively,
\be \label{torsions-from-multipliers}
\cT_{ij\perp}  = 0 \,,\quad \cT_{ij}{}^j = 0 \,.
\ee
Varying with respect to $\O_j{}^{ki}$ gives
\be \label{space_variation}
\cT_{ti\perp}\,\d^j{}_k - \cT_{tk\perp}\,\d^j{}_i + \e\,\lapse\,\cT_{il}{}^l\,\d^j{}_k - \e\,\lapse\,\cT_{kl}{}^l\,\d^j{}_i - 2\,\e\,\lapse\,\cT_{ik}{}^j - N^j\,\cT_{ik\perp} = 0 \,.
\ee
Finally, variation with respect to $\O_{[ij]\perp}$ yields
\be \label{antisymmetric-torsion}
-2\,\cT_{t[ij]} + \cT_{ki}{}^k N_j - \cT_{kj}{}^k N_i = 0 \,.
\ee
As already discussed, in the following we shall not need the variation with respect to $\O_{(ij)\perp}$ and therefore we refrain from exhibiting it. Still, eqs.~\eqref{torsions-from-multipliers}, \eqref{space_variation} and \eqref{antisymmetric-torsion} suffice to set to zero all components of the torsion that we included in eq.~\eqref{components-torsion}.
In particular, we are thus setting to zero the component $\cT_{ij}{}^{k}$ of the torsion.
By taking a linear combination of the cyclic permutations of this constraint over the indices $i$, $j$ and $k$ one obtains 
\be \label{space_vielbein_postulate_2}
e^k{}_A\,\pr_i\,e_j{}^A + \O_i{}^k{}_j - \g_{ij}{}^k = 0 \,,
\ee
where we introduced the Christoffel symbol $\g_{ij}{}^k$ of the Levi-Civita connection for the spatial metric.

\subsection{Rewriting of the action in Hamiltonian form}

Up to now, we proceeded without fixing any gauge for the local Lorentz frame. For simplicity, we now fix the time gauge, see eq.~\eqref{Carroll-frame}. Under this condition, $\text{n}^A=n^A = \d_0{}^A$, which is precisely eq.~\eqref{eq:internal_vector}, and the conjugate momenta to the spatial vielbein $\frame_i{}^a$ reads
\be
p^i{}_a \equiv \frac{\pr\cL}{\pr\dot{\frame}_i{}^a} = \frac{2\,\e^3}{16 \p G_N}\,\sqrt{g}\,\left(\O_k{}^i{}_\perp\,\frame^k{}_a - \O_k{}^k{}_\perp\,\frame^i{}_a \right) ,
\ee
where we used the time derivative of the second relation in eq.~\eqref{completeness} to transfer all time derivatives onto $\frame_i{}^a$. One can then rewrite, upon integration by parts, the first term in the action \eqref{NL:EHC} as
\begin{align}
\e^3\! \int \frac{dt\,d^D x}{8 \p G_N} \,\sqrt{g}\, \text{n}_{[B}\,e^i{}_{A]}\,\cR_{ti}{}^{AB} & = \int dt\,d^D x \,  p^i{}_a\, \dot{\frame}_i{}^a - \e^3 \! \int \frac{dt\,d^D x}{16 \p G_N} \sqrt{g} \left(\O_t{}^{ij} \cT_{ij\perp} - 2\,\O_t{}^i{}_\perp \cT_{ij}{}^j \right) \nn \\
& \approx \int dt\,d^D x \, \p^{ij}\, \dot{g}_{ij} \,,
\end{align}
where we used the symbol $\approx$ to stress that in the last line we imposed the constraints following from eqs.~\eqref{torsions-from-multipliers}, \eqref{space_variation} and \eqref{antisymmetric-torsion}. Moreover, we defined
\be\label{eq:piintermsofOmega}
\p^{ij} \equiv \frac12\,p^{(i}{}_a\,\frame^{j)}{}^a = \frac{\e^3\,\sqrt{g}}{16 \p G_N} \left(\O^{(ij)}{}_\perp - \O_k{}^k{}_\perp\,h^{ij} \right)
\ee
and we used that $\O_{[ij]\perp} = \text{n}_A\,\pr_{[i}\,e_{j]}{}^A = 0$ as a result of the first torsion constraint of eq.~\eqref{torsions-from-multipliers} and of the time-gauge condition $e_i{}^0 = 0$.

The term involving $\left( e^i{}_{[A}\,e^j{}_{B]}\,\cR_{ij}{}^{AB} - 2\, \Lambda \right)$ also splits into two parts. The first one constitutes the spatial curvature on account of $\frame_i{}^a\,\frame_j{}_a = g_{ij}$ and $\frame_i{}^a\,\frame^i{}^b = \d^{ab}$ and the spatial vielbein postulate \eqref{space_vielbein_postulate_2}:
\be
\frac{\e^4}{8\p G_N} \int dt\,d^D x\,\sqrt{g}\,\lapse \left[ e^i{}_{[A}\,e^j{}_{B]}\left(\pr_i\,\O_j{}^{AB} + \O_i{}^{AC}\,\O_j{}^{DB}\,\zeta_{CD} \right) - \Lambda \right] = - \int dt\,d^D x\,\lapse\,\cH_M,
\ee
where $\cH_M$ is the Hamiltonian constraint of magnetic Carrollian gravity defined in \eqref{eq:electricandmagneticH}. The second one can be rewritten as
\be
\frac{\e^4}{8\p G_N} \!\int\! dt\,d^D x\,\sqrt{g}\,\lapse\,e^i{}_{[A}\,e^j{}_{B]}\,\O_i{}^A{}_\perp\,\O_j{}^B{}_\perp{} =
\frac{16\p G_N}{\e^2} \!\int\! dt\,d^D x\,\frac{\lapse}{\sqrt{g}}\left( \pi^{ij}\,\pi_{ij} - \frac{1}{D-1}\,\pi^2 \right) ,
\ee
where we recognize the same structure as in the Hamiltonian constraint of electric Carrollian gravity defined again in \eqref{eq:electricandmagneticH}.

Finally, last term in eq.~\eqref{NL:EHC} reads
\be
\begin{split}
&\frac{\e^3}{8\p G_N} \int dt\,d^D x\,\sqrt{g} \left(N^i\,e^j{}_A\,\text{n}_B - N^j\,e^i{}_A\,\text{n}_B\right)\left(\pr_i\,\O_j{}^{AB} + \O_i{}^{AC}\,\O_j{}^{DB}\,\h_{CD} \right) \\
&\quad = 2 \int dt\,d^D x\,N_i\,\nabla_{\!j}\,\p^{ij} \,.
\end{split}
\ee

All in all, we arrive at
\be \label{ADM-action}
I = \int dt\,d^D x \left(\dot{g}_{ij}\,\p^{ij} - \lapse\,\cH_\perp - N_i\,\cH^i\right) ,
\ee
with
\begin{subequations}
\begin{align}
\cH_\perp &= \cH_M + \e^2 \frac{16 \p G_M}{\sqrt{g}}\,\left(g_{il}\,g_{jk} - \frac{1}{D-1}\,g_{ij}\,g_{kl}\right)\p^{ij}\,\p^{kl}  \,,\\[5pt]
\cH^i &= -2\,\nabla_{\!j}\,\p^{ij} \,,
\end{align}
\end{subequations}
which is exactly eq.~\eqref{eq:relativistic_ADM}, with the identification $G_N = \e^4 G_M$.

Notice that the combination of the surviving components of the spin connection giving $\pi^{ij}$ in \eqref{eq:piintermsofOmega} does not depend on $\e$ when written in terms of quantities that do not scale in the Carrollian limit:
\be
\pi^{ij} = - \frac{\sqrt{g}}{16\p G_M} \left(\omega^{(ij)} - h^{ij}\,\omega_k{}^k{} \right) ,
\ee
where only the spatial components of $\omega_\m{}^a = \omega_\m{}^{0a}$ appears. This implies, in particular, the relation
\be
S^{ab} = \frame^i{}_a\,\frame^j{}_b\,\omega_{(ij)}
\ee
which was derived in eq.~\eqref{sol-torsions}. Taking it into account in the limit $\e \to 0$ one recovers from \eqref{ADM-action} the action of magnetic Carrollian gravity in ADM form of eq.~\eqref{eq:magnetic-ADM}. Our present derivation of the relation between the action \eqref{ADM-action} and \eqref{eq:magnetic-ADM} highlights however how the Hamiltonian constraint of electric Carrollian gravity can be recovered as a subleading contribution starting from the first-order action.

\section{Conclusions}

In the gauging approach to Einstein's theory, the torsion-free, metric-compatible connection one-form plays a central role.  Since such a connection always exists in Riemannian geometry (and is unique), the gauging approach is well adapted to the description of the theory.

By contrast, the existence of a torsion-free, Carroll-compatible connection is the exception rather than the rule in Carrollian geometry \cite{Henneaux:1979vn, Vogel1965, Jankiewicz, Dautcourt1967}.   Such a connection exists only if the extrinsic curvature (or second fundamental form) vanishes, and in that case, it is not unique. It does not come as a surprise, therefore, that the gauging approach developed in  \cite{Bergshoeff:2017btm}, where a connection is introduced from the very beginning,  dynamically implements these features and is equivalent to the magnetic version of Carrollian gravity, for which exactly the same properties hold - the spatial symmetric tensor parametrizing the non-uniqueness of the connection being the conjugate momentum to the spatial metric.

In the electric version of Carrollian gravity, however, there is no torsion-free, Carroll-compatible connection since the extrinsic curvature does not vanish.  The standard gauging approach is for that reason not well adapted to the electric situation, since it assumes the existence of a physically relevant connection.  The non-existence of a natural connection sheds new light on the difficulties encountered in an orthodox gauging description of the Carrollian electric limit of Einstein's theory  \cite{Figueroa-OFarrill:2022mcy}. Further considerations on this problem are given in Appendix \ref{app:electric}.

Even though there is no natural connection, one could blame the problem on the choice of the Lagrangian.  In fact, one can write the second-order action of electric Carrollian gravity  \cite{Henneaux:1979vn} in terms of vielbein and connection as
\be \label{eq:electric_action_covariant}
I_{\hbox{\tiny{Elec}}}[\mathcal E_\mu{}^{A}, \omega_{\mu}{}^{AB}] \sim \int dt\,d^Dx \, \mathcal E \left( \Theta_{\la \mu} \Theta^{\la \mu} - \Theta^2 \right)
\ee
and argue that it fits already in the gauging framework, taking the form of a torsion-squared action.   Here, the symmetric, transverse tensor $\Theta_{\la \mu}$ is equal to the symmetrized components $e_\la{}^a  e_\mu{}^b T_{0(ab)}$ of the torsion, which we have seen does not involve the connection and is equal to the extrinsic curvature.  This is a bit artificial, however, since the connection does not appear in the Lagrangian and is thus a pure gauge field with no physical meaning - the gauge group contains more transformations than local Carroll transformations and diffeomorphisms and enables one to shift the connection at will.  This is as it should since there is no natural connection in the electric theory.\footnote{There is only a limited form of parallel transport involving the extrinsic curvature  \cite{Henneaux:1979vn}. Note 
that the action, which depends on the vielbein only through the metric and the volume element, can be written in covariant first-order form
$$
I_{\hbox{\tiny{Elec}}}[g_{\lambda \mu}, \Omega, P^{\lambda \mu}] = \int dt\,d^D x \, \Omega \Big[P^{\lambda \mu} K_{\lambda \mu} - G_{\lambda \mu \rho \sigma}  P^{\lambda \mu} P^{\rho \sigma}\Big]
$$
where $P^{\la \mu}$ is a symmetric tensor with gauge invariance $P^{\la \mu} \rightarrow P^{\la \mu} + \lambda ^\lambda n^\mu +  \lambda ^\mu n^\la$
and where
$
G_{\lambda \mu \rho \sigma} = \frac12\left(g_{\la \rho} g_{\mu \sigma} + g_{\la \sigma} g_{\mu \rho}\right) - \frac{1}{D-1} g_{\la \mu} g_{\rho \sigma} $. 
By eliminating $P^{\la \mu}$ through its own equation of motion, one recovers the second order action.}

\section*{Acknowledgements}

We wish to thank Arnaud Delfante, Joaquim Gomis and Stefan Prohazka for useful discussions. The work of A.C.\ and S.P.\ was supported by the Fonds de la Recherche Scientifique--FNRS under Grants No.\ FC.36447, F.4503.20 (HighSpinSymm) and T.0022.19 (Fundamental issues in extended gravitational theories). The work of M.H. was partially supported by the ERC Advanced Grant ``High-Spin-Grav'', by FNRS-Belgium (conventions FRFC PDRT.1025.14 and IISN 4.4503.15), as well as by funds from the Solvay Family. The research of A.P. is partially supported by Fondecyt grants No.\ 1211226 and 1220910. P.S-R. has received funding from the Norwegian Financial Mechanism 2014-2021 via the National Science Centre (NCN) POLS grant 2020/37/K/ST3/03390.

\appendix

\section{Comments on the electric limit}\label{app:electric}

In section~\ref{sec:limit} we saw how the Hamiltonian constraint of electric Carrollian gravity emerges from the first-order formulation, even if at a subleading order. It is thus natural to ask if there exists an alternative scaling of the components of the vielbein and of the spin connection allowing one to recover the full action of electric Carrollian gravity in ADM form. If one writes
\be \label{electric-rescaling}
E^\m{}_A = (\e^{-1}\,n^\m, \e^{-2}\,e^\m{}_a) \,,\quad E_\m{}^A = (\e\,\t_\m ,\e^2\,e_\m{}^a) \,,\quad \O_\m{}^{AB} = (\e\,\omega_\m{}^a, \e^2\,\omega_\m{}^{ab}) \,,
\ee
which differs from eq.~\eqref{rescalings} by the addition of a factor of $\e^2$ (resp. $\e^{-2}$) on $e_\m{}^a$ and $\omega_\m{}^{ab}$ (resp. $e^\m{}_a$), chooses the time gauge once again so that the metric field takes the expression $g_{ij} = \e^4\,\frame_i{}^a\,\frame_j{}_a$, and also rescales Newton's constant as
\be \label{electric-GN}
G_N = \e^{2D+2}\,G_E \,,
\ee
then, the relativistic action takes again the ADM form \eqref{ADM-action}. This time, however, the Hamiltonian density takes the form
\begin{subequations}
\begin{align}
\cH_\perp &= \cH_E
- \e^2 \frac{\sqrt{g}}{16 \p G_E}\left(R - 2\La \right) \,,\\
\cH^i &= - 2\,\nabla_{\!j}\,\p^{ij} \,,
\end{align}
\end{subequations}
with
\be
\p^{ij} = - \frac{\sqrt{g}}{16\p G_E}\, \omega_k{}^a \left(\frame^{(i}{}_a\,h^{j)k} - \frame^k{}_a\,h^{ij} \right) = - \frac{\sqrt{g}}{16\p G_E}\left(\omega^{(ij)} - h^{ij}\,\omega_k{}^k \right) .
\ee
When $\e \to 0$, we recover the electric theory
\be \label{eq:electric_action}
I_E=\int dt\,d^{D}x \left(\pi^{ij}\dot{h}_{ij}-\lapse \mathcal{H}_{E}-N^{i}\mathcal{H}_{i}\right) ,
\ee
which is equivalent to eq.~\eqref{eq:electric_action_covariant} after elimination of the conjugate momenta. One can also rescale the cosmological constant $\La = \e^{-2}\,\La_E$ to obtain a non-zero cosmological constant term in the limit in agreement with \cite{Hansen:2021fxi}.

There is an important proviso though: if one implements the limit $\e \to 0$ in the Einstein-Cartan action with the rescalings \eqref{electric-rescaling} and \eqref{electric-GN} one obtains the action 
\be \label{wrong-limit}
I_E = \frac{1}{16\p G_E}\int dt\,d^D x\, \cE \left(2\,e^\m{}_a\,e^\n{}_b\left(\pr_{[\m} \, \omega_{\n]}{}^{ab} + \omega_{[\m}{}^{a} \, \omega_{\n]}{}^{b} \right) + 4\,n^\m\,e^\n{}_a\,\pr_{[\m}\,\omega_{\n]}{}^{a} \right) .
\ee
This action is not invariant neither under local Carrollian boost nor under local spatial rotations. Choosing the time gauge introduces therefore a non-trivial restriction in this context. Also, the rescaling \eqref{electric-rescaling} affects the definition  \eqref{spatial_metric} of the spatial metric as
\be
g_{ij} = \e^4\,e_i{}^a\,e_j{}^b\,\d_{ab} - \e^2\,e_i{}^0\,e_j{}^0 \,.
\ee
Its $\e \to 0$ limit is very different whether one chooses the time gauge or not, since the matrix $e_i{}^a\,e_j{}^b\,\d_{ab}$ is assumed to be invertible, but $e_i{}^0\,e_j{}^0$ is only of rank $1$. This has to be contrasted with the magnetic Carrollian limit of section \ref{sec:limit}, where the choice of the time gauge did not have any effect on the form of the spatial metric in the $\e \to 0$ limit.

Those observations are in line with the results of \cite{Figueroa-OFarrill:2022mcy}, where it was shown that it is not possible to recover the electric Carrollian theory by gauging the Carroll algebra.


\end{document}